\documentstyle[12pt]{article}
\begin{document}
\setlength{\baselineskip}{.7cm}
\renewcommand{\thefootnote}{\fnsymbol{footnote}}
\sloppy

\begin{center}
\centering{\bf RANK-ORDERING STATISTICS OF EXTREME EVENTS:
APPLICATION TO THE DISTRIBUTION OF LARGE EARTHQUAKES}
\end{center}

\begin{center}
\centering{\bf Didier Sornette$^{1}$, Leon Knopoff$^{2,3}$, Yan
Kagan$^{3}$ and Christian Vanneste$^{1}$} \\
\end{center}

\begin {center}
\centering{$^{1}$ Laboratoire de Physique de la Mati\`ere Condens\'ee,
CNRS URA 190\\ Universit\'e de Nice-Sophia Antipolis, B.P. 71\\ Parc
Valrose, 06108 Nice Cedex 2, France\\
$^{2}$ Department of Physics, University of California
Los Angeles, California 90095 and\\
$^{3}$ Institute of Geophysics and Planetary
Physics\\ University of California, Los Angeles, California 90095-1567}
\end{center}

\renewcommand{\thefootnote}{\fnsymbol{footnote}}

{\bf Abstract} \\

Rank-ordering statistics provides a perspective on the rare, largest
elements of a population, whereas the statistics of cumulative
distributions are dominated by the more numerous small events. The
exponent of a power law distribution can be determined with good
accuracy by rank-ordering statistics from the observation of only a
few tens of the largest events. Using analytical results and synthetic
tests, we quantify the systematic and the random errors.

We also study the case of a distribution defined by two branches, each
having a power law distribution, one defined for the largest events
and the other for smaller events,  with application to the World-Wide
(Harvard) and Southern California earthquake catalogs. In the case of
the Harvard moment catalog, we make more precise earlier claims of the
existence of a transition  of the earthquake magnitude distribution
between small and large earthquakes; the $b$-values are $b_2 = 2.3 \pm 0.3$ for
large shallow earthquakes and $b_1 = 1.00
\pm 0.02$ for smaller shallow earthquakes. However, the cross-over magnitude
between the two distributions is ill-defined. The data available at
present do not provide a strong constraint  on the cross-over which
has a $50\%$ probability of being between magnitudes $7.1$ and $7.6$
for shallow earthquakes; this interval
may be too conservatively estimated. Thus, any
influence of a universal geometry of rupture on the distribution of
earthquakes world-wide is ill-defined at best. We caution that there
is no direct evidence to confirm the hypothesis that the large-moment
branch is indeed a power law.  In fact, a gamma distribution fits the
entire suite of earthquake moments from the smallest to the largest
satisfactorily. There is no evidence that the earthquakes of the
Southern California catalog have a distribution with two branches, or
that a rolloff in the distribution is needed; for this catalog $b =
1.00 \pm 0.02$ up to the largest magnitude observed, $M_W \simeq 7.5$;
hence we conclude that the thickness of the seismogenic layer has no
observable
influence whatsoever on the frequency distribution in this region.

\pagebreak

\section{Introduction}

The observation that many natural phenomena have size distributions
that are power laws, has been taken as a fundamental indication of an
underlying self-similarity [{\it Mandelbrot}, 1983; {\it Aharony and
Feder}, 1989;
{\it Riste and Sherrington}, 1991]. A power law distribution indicates the
absence
of a characteristic size and as a consequence that there is no upper limit on
the size
of events. The largest events of a power law distribution completely
dominate the
underlying physical process; for instance, fluid-driven erosion is dominated by
the largest floods and most deformation at plate boundaries takes place through
the agency of the largest earthquakes. It is a matter of debate whether
power law distributions, which are valid descriptions of the numerous
small and intermediate events, can be extrapolated to large events;
the largest events are, almost by definition, undersampled.  Based on
analytical calculations and synthetic tests, we demonstrate that the
observation of a few tens of the largest events is sufficient to
qualify their distribution with good precision, using the
rank-ordering technique, initially introduced in linguistics [{\it Zipf}, 1949]
and
later largely used in statistics [{\it Epstein and Sobel},
1953; {\it Gumbel}, 1960; {\it Deemer and Votaw}, 1955].

In the case of earthquakes, a number
of authors have proposed a geometrical argument that the
distribution for large earthquakes should be substantially different
from that for small ones because of the one-dimensional character of
the fracture surface of large earthquakes and its two-dimensional
character for small earthquakes [{\it Kanamori and Anderson}, 1975; {\it
Geller},
1976; {\it Shimazaki}, 1986; {\it Rundle}, 1989; {\it Scholz}, 1990;
{\it Romanowicz},
1992; {\it Pacheco et al.}, 1992; {\it Romanowicz and Rundle},
1993; {\it Okal and Romanowicz}, 1994].  More specifically
they propose that the large-magnitude branch of the distribution is also a
power
law and that the cross-over moment or magnitude between these two distributions
is a measure of the thickness of the seismogenic zone in the case of
strike-slip
earthquakes and of the downdip dimension of rupture in the case of
earthquakes in
subduction zones. {\it Pacheco et al.}\ [1992] claim to have identified a
kink in the
distribution of shallow transform fault earthquakes around magnitude
5.9 to 6.0, which corresponds to a characteristic dimension of about
10km;   a kink for subduction zones is presumed to occur at a moment
magnitude near 7.5, which corresponds to a down-dip dimension of the
order of 60km.

It is possible to show that the power-law
distribution that holds for small earthquakes cannot be extended to
infinity because it would imply that an infinite amount of energy be
released from the earth's interior [{\it Knopoff and Kagan},
1977; {\it Kagan and Knopoff}, 1984; {\it Kagan}, 1994].
According to either the
energy or the geometrical arguments, there must be a cross-over or a
rollover to a second branch of the distribution.  According to the
energy argument, a truncation of the distribution for small earthquakes
is not forbidden; there is however the following intuitive reason why a cutoff
should not be abrupt: we see no reason why the distribution should be
finite for a magnitude of let us say 7.99, and zero for magnitude 8.00
[{\it Kagan and Knopoff}, 1984].
Although there is no compelling argument
that the branch of the distribution for earthquakes with large
energies must be a power law, it is clear that if this distribution is
indeed a power law, then it must have an exponent that is larger than
that for smaller earthquakes; specifically, the energy in the power
law cumulative distribution of large earthquakes should fall off as
$E^{-b/\beta}$ with $b/\beta > 1$
[{\it Knopoff and Kagan}, 1977], where $b$ is the usual exponent in
the magnitude-frequency law and $\beta$ is generally taken to be
$1.5$.

We analyze the problem of the statistics of power law distributions in
detail.  We pay special attention to the case of a distribution
composed of two power laws, one for the extreme events and the other
for intermediate and small events. We find that a modest amount of
data for the largest events is sufficient to determine the exponent
for this branch, whereas the value of the cross-over is ill-defined,
unless one has data for a much larger number of large magnitude
events.  We apply this technique to the analysis of the Harvard catalog
[{\it Dziewonski et al.}, 1993] of world-wide earthquake moments and to the
Southern California earthquake catalog [{\it Hutton and Jones}, 1993].

Consider a stochastic process in which the outcome, which we call the
energy $E$ generically, has the distribution

$$P(E) dE = \frac{C}{E^{1+\mu}}  dE ,  \eqno{(1)}$$
\\
with $E_{min} = 1 \leq E < \infty$; then $C= \mu$. Suppose that $N$
events occur within a given time interval. Let $E_1 \geq E_2 \geq ...
\geq E_n \geq... \geq E_N$ be the energies of the events listed in
descending order. If $N$ is large, the power law distribution (1) is
easily retrieved by the standard procedure of constructing a log-log
plot of the number of events in a given energy interval. Suppose
however that $N$ is small, let us say of the order of $10$ or at most
$100$. In this case, the number of events will be much too small to
permit the determination of $P(E)$ by standard techniques.

For illustrative purposes, consider the $100$ largest events to be
found in a random selection of $N= 10^5$ samples from the power law
distribution (1) with exponent $\mu = 1$. We thus mimic the standard
situation where only a fraction of the total population can be
observed due to certain limitations of resolution. Our selection
procedure has the same statistics as one in which we use $N=100$,
$E_{min} = 1000$ and keep all $100$ events generated. We plot $E_n$ as
a function of $n$ (Fig. 1) for five realizations of the random
process.

Figure 1 allows us to quantify our claim that the power law
distribution can be extracted with reasonable precision from a few
tens of the largest samples. A least-squares fit to each of these five
realizations in the interval $1 \leq n \leq 100$ shows that $E_n$
scales as $E_n \sim n^{-\gamma}$ with an exponent ranging from  $0.85$
to $1.22$. We can show that $\gamma = \frac{1}{\mu}$, which is easily
obtained from the equation $N \int_{E_n}^{+\infty} P(E) dE  \simeq
n$; the latter constraint requires that on average there be $n$ values
of $E$ larger than or equal to $E_n$ out of $N$ samples. Our estimate
for $\mu$ thus ranges from $0.83$ to $1.19$, which should be compared
with the exact value $\mu=1$. If we increase the value of $N$, the
precision of the determination of the exponent $\mu$ improves rapidly;
if we increase $N$ to $N =10^5$, we get
$\gamma = \mu = 1.00 \pm 0.01$.

The rank-ordering technique is fundamentally different from attempts
to improve the determination of the $b$-value from extreme-value
statistics. This latter technique involves a discarding of a
significant part of the data, and has been shown to be
unreliable [{\it Knopoff and Kagan}, 1977]. The rank-ordering technique
makes full
use of all of the data available; we analyze it by studying the largest events.

The rank-order distribution is the same as the usual cumulative
distribution but with an interchange of axes.  However the statistical
analyses of the two seemingly equivalent versions are significantly
different:  the statistics of the rank-ordering procedure focuses on
the values of the energies or moments of the samples; the statistics
of the cumulative distributions is concerned with uncertainties of the
order numbers, which are integers. Rank-ordering statistics therefore
provides a different perspective, with emphasis on the extreme tail of
the distribution. We show that the rank-ordering statistics ensures
that the fits are constrained by the largest events and not by the
most numerous smaller ones.

\section{One power law: Analytical results}

Before proceeding with numerical tests of the rank-ordering method, we
state the following results. The probability $F(E_n) dE_n$ that the
energy for the nth order event be equal to $E_n$ within $dE_n$
is [{\it Gumbel}, 1960] :

$$F(E_n) dE_n = {N\choose n}
 \biggl(1 - \int_{E_n}^{+\infty} P(E) dE \biggl)^{N-n} {n\choose 1}
 \biggl(\int_{E_n}^{+\infty}  P(E) dE \biggl)^{n-1}  P(E_n) dE_n
\eqno{(2)}$$
\\
for an arbitrary distribution $P(E)$. For the power law, $F(E_n)$ has
a peak at $E_n^{max}=[\frac{(\mu N + 1)}{\mu n + 1}]^{1/\mu}$, which
recovers and makes precise the scaling law above. If $P(E)$ is given
by the power law (1), $F(E_n)$ can be expanded around this maximum as
$$F(E_n) = F(E_n^{max}) -
\frac{1}{2} |\frac{d^2 F(E_n)}{dE_n^2}|_{E_n^{max}} (E_n - E_n^{max})^2 +
...,$$
\\
which allows us to get an estimate of the standard deviation of $E_n$ through
the calculation of $\Delta E_n \equiv <(E_n -
E_n^{max})^2>^{\frac{1}{2}}$, where the brackets indicate that the
statistical average is taken. We obtain

$$\frac{\Delta E_n}{E_n^{max}} = (\frac{1}{\mu(n\mu +1)})^{\frac{1}{2}} .
\eqno{(3)}$$
\\
We have used this expression to construct the two straight lines in
Fig. 1 that represent one standard deviation from the best fit. We
recover the usual general result, also observed in Fig. 1, that
$\frac{\Delta E_n}{<E_n>}$ decays as  $\frac{1}{\surd n}$ for $n\gg
1$.

A maximum likelihood estimate for $\mu$ can be derived
directly [{\it Aki}, 1965; {\it Deemer and Votaw}, 1955; {\it Hill}, 1975].
Consider a variable $x$
that has the normalized exponential distribution $P(x) = \mu e^{-\mu x}$,
$0\leq
x<\infty$. Suppose that we measure only values of $x$ larger than some $x_0$
and
that the total number of trials $N$ is not known. Then the probability
distribution, restricted to have values larger than $x_0$, is
$P_{\mu}(x) = \mu e^{-\mu(x-x_0)}$. For a set of $n$ samples ($x_1$,
$x_2$, ..., $x_n$)  each greater than $x_0$ and having average value
$\bar{x}$, the probability for this
set is simply $P_{\mu}^n(x) = P_{\mu}(x_1)... P_{\mu}(x_n)$, which can
be written as $P_{\mu}^n(x) = \mu^n e^{-\mu n(\bar{x}-x_0)}$.
The maximum likelihood estimate of the most probable value
$\mu^*$ of $\mu$, is given by $\frac{dP_{\mu}^n}{d\mu}|_{\mu^*} = 0$,
and is $\mu^* = \frac{1}{\bar{x}-x_0}$. Using
rank-ordering for ranks $1$  to $n$ thus yields the estimate $\mu =
\frac{1}{\frac{1}{n} \sum_{i=1}^n x_i-x_n}$, where $x_n$ is the $n$th
largest value. Under the change of variable $E=e^x$,  $E$ is
distributed according to a power law (1) with exponent $\mu$ if $x$ is
exponentially distributed. Given the first $n$ largest values $E_1
\geq E_2 \geq ...\geq E_n$, the maximum likelihood estimate for the
exponent $\mu$ is simply the maximum likelihood estimate of $\mu$ of
the exponential distribution,  expressed in terms of the variable $E$,
since, if the probability is maximum for the exponential case, it is
also maximum for $\mu$ with this monotonic increasing change of variable. Thus
the maximum likelihood estimate for $\mu$ is
[{\it Aki}, 1965; {\it Deemer and Votaw}, 1955; {\it Hill}, 1975]

$$\mu = \frac{1}{\frac{1}{n} \sum_{i=1}^n \log \frac{E_i}{E_n}} . \eqno(4)$$
\\
The accuracy of this estimate is given by the standard deviation
$\Delta \mu = \biggl( \frac{d^2 \log P_{\mu}^n}{d\mu^2}
\biggl)^{\frac{1}{2}}$ which yields $\frac{\Delta \mu}{\mu} =
\frac{1}{n^{1/2}}$, showing that the accuracy of the determination of
$\mu$ improves  according to the usual law of large numbers.

We can derive the full distribution of $\mu$. Let $\mu_0$ be the true
value of the exponent. In order to derive the distribution of $\mu$
from a finite data sample, we note that $$<\log E> |_{E>E^*} =
\int_{E^*}^{+\infty} \log E  P(E) dE = \log E^* + \frac{1}{\mu_0}.$$
The quantity $<x>|_{E>E^*}$ is the average of $x$ with respect to its
distribution, for those samples with $E$ conditioned to values larger
than $E^*$.  With eq.\ (4) which can also be written
$$<\frac{1}{\mu}>|_{  E>E_n} = <\log E> |_{E>E_n} - \log E_n,$$ this
yields $$<\frac{1}{\mu}>|_{E>E_n} = \frac{1}{\mu_0}.$$  We next note
that $$<(\log E)^2> |_{E>E^*} = (\log E^*)^2 + \frac{2 \log
E^*}{\mu_0} + \frac{2}{\mu_0^2},$$ which yields $$Var[\log E] = <(\log
E)^2> |_{E>E^*} - <\log E> ^2 |_{E>E^*} = \frac{1}{\mu_0^2}.$$ With
eq.\ (4), we obtain $$Var[\frac{1}{\mu}] = Var[\frac{\sum_{i=1}^n \log
E_i - n \log E_n}{n}] = \frac{1}{n\mu_0^2}.$$ The central limit
theorem states that the distribution of $\sum_{i=1}^n \log E_i$ will
have an approximate normal distribution for sufficiently large $n$.
Thus, for known $\mu_0$, $\frac{1}{\mu}$ has an approximately normal
distribution of mean $\frac{1}{\mu_0}$ and variance
$\frac{1}{n\mu_0^2}$ : $$P(\frac{1}{\mu}) = (\frac{n}{2\pi})^{1/2}
\mu_0 e^{-(\frac{1}{\mu} - \frac{1}{\mu_0})^2 n\mu_0^2/2}.$$ The
distribution $P(\mu)$ for $\mu$ is simply deduced by using the formula
$P(\mu) = P(\frac{1}{\mu}) \frac{d(1/\mu)}{d\mu}$, yielding

$$P(\mu) = (\frac{n}{2\pi})^{1/2} \frac{\mu_0}{\mu^2}
 e^{-\frac{n(\mu-\mu_0)^2}{2\mu^2}}. \eqno(5)$$
\\
We note that the distribution $P(\mu)$ is skewed due to the prefactor
$\frac{\mu_0}{\mu^2}$ in front of the exponential term. We discuss its
practical importance below.

It is
useful to give the maximum likelihood estimator for $\mu$ for the case
where the data values are restricted in a finite range
$E_s \leq E \leq E_L$. The previous case corresponds to $E_L \rightarrow
+ \infty$ and $E_s = E_n$. In the present case, we
consider the possibility that,
in some applications, the estimates of the
higher values of magnitude
may be unreliable, for instance their
recordings may saturate the seismograph, or
these estimates may be
biased by finite size effects as may occur in some
numerical simulations. In these
cases, expression (4) should not be used but rather a modified version
which takes into account the fact that values larger than some
$E_L$ are not taken into account. A procedure that could be used
in future work would be to examine the dependence of $\mu$
as a function of $E_L$. We start as above with
the expression of the probability distribution for the exponential
variable $x$, restricted to have values larger than $x_s$:
$P_{\mu}(x) = \mu e^{-\mu(x-x_s)}$, $x_s\leq x<\infty$. Using
the identity $P(A/ {\rm knowing} B) = {P(A) \over P(B)}$, the distribution
restricted to have values $x$ between $x_s$ and $x_L$ is then simply
$$P_{\mu}(x) = \frac{\mu e^{-\mu (x-x_s)}}{1- e^{-\mu x_L}} .$$
$x_1$, $x_2$, ..., $x_n$ are all between $x_s$ and $x_L$ with probability
$$P_{\mu}(x_1)...P_{\mu}(x_n) =
\frac{\mu ^n e^{-n \mu (\bar{x}-x_s)}}{\biggl( 1-  e^{-\mu x_L} \biggl)^n} .$$
The maximum likelihood value for $\mu$ then obeys
$${1 \over \mu} - (\bar{x}-x_s) - {1 \over n} \frac{x_L e^{- \mu x_L}}{1- e^{-
\mu x_L}}  = 0 ,$$
which gives the previous result for $x_L \rightarrow + \infty$.
With the change of variable $E=e^x$, we get the maximum likelihood
estimate for the exponent $\mu$ from the knowledge of all
the data values between $E_s$ and $E_L$ :
$${1 \over \mu} - \frac{1}{n} \sum_{i=1}^n \log \frac{E_i}{E_s}
 - {1 \over n} \frac{E_L^{-\mu} \log
E_L}{1- E_L^{-\mu}}  = 0 ,$$
which recovers eq.\ (4) for $E_L \rightarrow + \infty$
{\it Deemer and Votaw}, 1955].

These results highlight the power of the rank-ordering technique and
its generalization to
extract the power law distribution from the information contained in
the largest events, which characterize the large-event tail of the
distribution.  The largest events constrain the tail of the
distribution drastically and thus permit a surprisingly good recovery
of the exponent from a relatively small data set. It is clear that the
statistics improves if the number of the very largest events we
consider increases, as shown by (3) and Fig. 1.

\section{Two power laws: Synthetic tests}

In view of our application of rank-ordering methods to earthquake
data, we present tests of the method for synthetic distributions
constructed from two power laws, each valid over a different range of
energy. Let  $P(E) = \frac{C_1}{E^{1+\mu_1}}$ for $1 \leq E \leq E_c$
and $P(E) = \frac{C_2}{E^{1+\mu_2}}$ for $E_c \leq E < \infty$. The
parameters $C_1$ and $C_2$ are determined from the normalization of
$P(E)$ and the condition of continuity at $E_c$, for given $\mu_1$,
$\mu_2$ and the cross-over value $E_c$. For a given number of events
$N$, there is an exact relationship between the cross-over value $E_c$
and the average number of events $n_2$ that sample the power law for
the larger events with $E > E_c$:

$$E_c = (1 +
\frac{\mu_1}{\mu_2} (\frac{N}{n_2} -1))^{\frac{1}{\mu_1}}. \eqno(6) $$.
\\
If the number of great events $n_2$ is large, corresponding to a
not-too-large $E_c$, then the two power laws can be reconstructed
easily through the standard log-log plot of the number of events in a
given energy interval. However this will not be the case in the
seismological example below. Therefore we consider the more difficult
case where the number of large events $n_2$ is small, let us say
of the order of a few tens of events.

Figure 2 shows the rank-ordering for a single realization with
$N = 10^5$, $\mu_1 = \frac{2}{3}$, $\mu_2 = \frac{4}{3}$ and $n_2 =
50$, corresponding to $E_c \sim 10^5$. (These choices for the
exponents $\mu_{1,2}$ were made in view of values suggested in
{\it Pacheco, et al.}\ [1992].)
The two branches are clearly identifiable.
The accuracy with which the exponent for the lower energy branch
$\mu_1$ can be determined is of course excellent since the statistics
covers more than
$3$ decades.
However, since only about $50$ extreme events sample
the power law branch for large energies,
there are large fluctuations in this
branch, which
raises the question whether there is
a possibility to extract the exponent of the second power law and the
cross-over energy from an analysis of the first tens of events with the
largest energies in the extreme tail of the distribution. To determine
the statistical accuracy of the value of the exponent $\mu_2$ for the
largest events, we have generated $10^5$ different realizations with
the parameters used in Fig. 2 and listed above. For each realization
we select the $p$ largest events with $p= 25, 50, 75, 100$ and $200$.
For each $p$, we use eq.\ (4) to get an estimate of $\mu_2$ with $n$
running from $5$ to $p$ to establish the stability of the
determination of  $\mu_2$. We have also  made a least-square fit of
the data $E_n$ with a power law over the  $p$ points. In practice for
this limited range $25 \leq p \leq 200$, the two methods give results
which are essentially indistinguishable. However, the maximum
likelihood estimate of eq.\ (4) is more sensitive to deviations from a pure
power law. Although $n_2$ was fixed at $50$ in the realizations, we
assume that this number is not known in the analysis; thus we analyze
the data as though we had no {\it a priori} information on the value
of the cross-over, since this would not be generally known in
earthquake data. The histogram of the number of realizations that have
a given value of $\mu_{2,est}$ is shown in Fig. 3. The distribution of
$\mu_{2,est}$ is significantly skewed and peaks at a value below the
true value of $4/3$. The maximum value of $\mu_{2,est}$ increases
progressively toward $4/3$ as $p$ increases up to around $p = 75$
where it is close to $1.2$ and then decreases for larger values of
$p$; for $p=200$, the maximum is around $1.1$. The latter decrease is
due to the fact that the small-energy branch begins to influence
significantly the determination of the slope of the large-energy
branch if too many samples with rank orders $n > n_2$ are included, as can be
observed directly from Fig. 2. The observed skewness and shift of the
maximum are predicted from eq.\ (5), since the maximum of $P(\mu)$
occurs at $\mu = \mu_0 \frac{2}{1 + (1+[8/p])^{1/2}}$, a value
that is always smaller than the true value $\mu_0$. For example, if
$p=25$, then $\mu = 0.93 \mu_0$ and thus $\mu = 1.24$, for $\mu_0 =
4/3$. Although this remark makes the existence of skewness plausible,
the actual skewness is larger than predicted by eq.\ (5), which we
attribute to the fact that equation (5) does not allow for additional
fluctuations and distortions of the statistics due to the presence of
the two power law branches in the distribution.

The curves in Fig. 3 are reminiscent of log-normal or Weibull
distributions, being characterized by a sharp decay on one side and a
long tail on the other side of the peak. Consider the Weibull
distribution,

$$p(\mu) = C m \mu^{m-1} e^{-C\mu^m}. \eqno(7)$$
\\
We can easily derive the two parameters $C$ and $m$ from the {\it a
priori} knowledge of $<\mu> =\frac{4}{3}$ and from the direct
observation of the most probable value $\mu_{x}$ which is the
value at the maximum of the histogram of Fig. 3. We note in
particular that the ratio $\frac{<\mu>}{\mu_{x}}$ is a function
solely of $m$. Thus we have a simple and direct determination of $m$
for each case $n= 25, 50, 75, 100, 200$, which then allows us to estimate
the width of the distribution, defined by the square root of its
variance $V$. For instance, for $n=50$, we measure $\mu_{2} \simeq
1.08$, leading to $\frac{<\mu>}{\mu_{2}}\simeq 1.25$. This is
exactly the value expected for $m=2$ for which we compute
$V^{\frac{1}{2}} = 0.31$, in good agreement with inspection of Fig. 3.
Thus for $n=50$, we have $\mu_{2,est} = 1.1 \pm 0.3$, which should be
compared with the exact value $\mu_2=1.33$. For $n=75$ and $100$, we
get $\mu_{2,est} = 1.15 \pm 0.25$ ($m\simeq 2.4$). Thus we have
quantified the skewness of the distribution: for our choice of
parameters, the most probable measured value of
$\{\mu_{2,est}\}_{x}$ is about $0.2$ below the true value $1.33$,
with a root-mean-square error of about $0.3$.

These results complement the simple maximum likelihood estimate
derived above which states that the determination of $\mu_2$ has a
relative error that is given by $\frac{\Delta \mu}{\mu} =
\frac{1}{n^{1/2}}$. For $n=50$ and $\mu_2=1.33$, this yields $\Delta
\mu \simeq 0.19$, which is smaller than our numerical estimate of
$0.3$. The synthetic tests have shown that there is a notable skewness
of the same order of magnitude as the error. These results concerning
the standard deviation and skewness are relevant to any interpretation of
observations of earthquake data.

We also observe that, as $n$ becomes smaller, the probability that we
measure a large value of $\mu_{2,est}$ becomes larger. For example, if
$n=5$, the tail extends to $\mu_{2,est}=5$, and in about $5-10 \% $ of
the realizations we find values of $\mu_{2,est}$ greater than $2$.
This reflects the large fluctuations in the energies for the lowest
rank-order events which are those with the largest energies, as can be
seen from (3). Using $\mu=4/3$, we find  $\frac{\Delta E_1}{E_1^{x}}
\simeq 0.66$ from (3), for the event with the largest energy.

We have also studied the properties of the rank-ordering technique for
different values of $\frac{\mu_2}{\mu_1}$ and different $n_2$. We find
that the larger the value of $n_2$, the easier and more precise the
determination of $\mu_2$. However, for very small $n_2 \sim 10$, there
are some practical limitations. For instance, for $\mu_1 =
\frac{2}{3}$ and $\mu_2 = 1$, one can identify the presence of a
cross-over, but the exponent $\mu_2$ that we derive is hugely
overestimated by as much as a factor of $2$. On the other hand, if
$\frac{\mu_2}{\mu_1}$ is of the order of or larger than $2$, the
bias in the determination of $\mu_2$ approaches that of the
determination  of the exponent for a distribution without two
branches, i.e. as though the distribution was that of a single power
law with the exponent $\mu_2$, as measured from the few tens of events
with the largest energies, as discussed above.

\section{Cross-over energies}

We consider the question of the determination of the cross-over values
$n_2$ and $E_c$. Since we would like to understand the physical
process responsible for the cross-over from the measurement of the
cross-over value, in principle we should use a statistical method of
interpretation which is independent of any {\it a priori} assumptions.
This is a difficult problem, since the functional form of the
cross-over is not known in general. It is clear from Fig. 2, for
example, that the cross-over is not sharp, simply because of the
cumulative nature of the rank-ordering procedure.

We have attempted to estimate $n_2$ by three different techniques.
First, it is natural to use eq.4 and plot $\mu$ as a function of $n$.
For small $n$, one observes large fluctuations as expected.  But
typically, as soon as $n$ is of the order of $10$, $\mu$ enters a regime
where it fluctuates around a constant average value with rough
relative amplitudes $\frac{1}{\sqrt n}$. We might hope to detect the
cross-over $n_2$ as corresponding to the rank $n^*$ beyond which a
significant deviation from this constancy is observed. In those cases
in which we did detect a bend or deviation from a constant value, we
found that the corresponding rank $n^*$ had little to do with the
average expected value $n_2=50$. Indeed, we observed $n^*$ to
fluctuate typically between $20$ or even less and about $100$.

A second attempt to estimate $n_2$ was made using the least-square fit
method. We plotted the cost function, i.e. the rms of the residuals in
the least-squares fit to the first $n$ rank-ordered samples,
normalized by the number of events $n$.  Our hope was that,
for increasing $n<n_2$, the cost function would show a tendency to
decrease as the large-energy branch of the distribution with exponent
$\mu_2$ is approximated better and better. However, for  $n$ larger
than the true cross-over value $n_2$, the cost function would start to
increase, and as $n$ becomes significantly larger than $n_2$, the fit
to the overwhelming number of events in the smaller-energy branch
should yield a single power law that would be insufficient to account
for the cross-over between the two power laws; hence the cost function
should deteriorate significantly. Thus we expect
$n_2$ to be the value that minimizes the chi-square estimate.
Among $1000$ realizations with the
same generating parameters listed above, we did not find any
consistent behavior that would enable us to determine the cross-over
value with good accuracy. Indeed, members of one subset of these of
realizations have only a weak and often broad minimum in the cost
function at values that differ from realization to realization in the
range $n = 5$ to $50$; most realizations do not display any minimum
whatsoever below $n = 100$. In all cases, the cost function is
extremely variable from realization to realization. We are forced to
conclude that it is not possible to get a good estimate of $n_2$ by
this method either.

In our third attempt to estimate the crossover, we
used a less sophisticated method, namely we took the best estimate of
$n_2$ to be the rank value at which the large-energy branch of the
log-log rank-ordering plot begins to bend over by inspection.  Because of the
cumulative nature of the rank-ordering procedure, this definition is
relatively precise; what is required is to identify the bendover.
Other estimates of the cross-over, such as the intersection between
the extrapolations of the two linear parts of the distribution are
clearly biased. We have found that the estimate from the onset of
bendover yields values which fluctuate between $n = 10$ to $100$ or
more, while the value from which the data were synthesized is
$n_2=50$. This wide range of estimates arises despite the existence of
an apparent well-defined bend in the plots.

Thus the appearance of a bend or kink on rank-order or cumulative
plots should be interpreted with caution. We are led to suggest a
warning:  a given realization may exhibit an apparent cross-over at a
rank-order value which has a large probability of being far from the true value
by as much as a factor of two! Quantitatively, only about half the
realizations have their apparent cross-over between $30$ and $100$
distributed around the true value of $50$. Below, we comment on the
relevance of these results to the earthquake problem.

\section{Application to earthquake catalogs}

We apply the rank-ordering technique to large earthquake catalogs. We
address the problem of the determination of the Gutenberg-Richter
magnitude-frequency law, which is often written as the cumulative
distribution

$$\log_{10} N_> = a - b M_W; \eqno(8) $$
\\
$N_>$ is the number of earthquakes whose magnitudes are equal or
greater than $M_W$, where $M_W$ is the magnitude, defined by

$$M_W = \frac{1}{\beta} [\log_{10} (m_0) - 9 ],  \eqno(9) $$
\\
where $m_0$ is the seismic moment in N.m; $\beta$ is generally taken
equal to $1.5$. If we combine these two expressions, we get a power
law distribution for the number of earthquakes having a given seismic
moment that is identical to equation (1), characterized by the
exponent $\mu = \frac{b}{\beta}$.  For small and intermediate
magnitude earthquakes, $b\approx 1.0$; thus $\frac{b}{\beta} \approx
2/3$.

We revisit the Harvard and the Southern California
catalogs. The Harvard catalog we use [{\it Dziewonski et al.}, 1993]
spans the time interval from $1977$ to $1992$ for earthquakes
worldwide, and the Southern California catalog, 1932-1991 [{\it Hutton and
Jones}, 1993]
spans almost $60$ years
of local seismicity. Both catalogs, as well as all other catalogs that
span a broad range of magnitudes reliably, are dominated by the large
numbers of small earthquakes; the time span of these catalogs is
significantly shorter than the time interval between the strongest
earthquakes, and as a consequence the data for the strongest
earthquakes is sketchy at best. A search for a kink in the
magnitude-frequency distribution has several hazards, the most obvious
of which is the saturation in the magnitude scale, which taken
literally, automatically generates a bend, if not a kink.
{\it Kanamori} [1977] has used the seismic moment to derive a magnitude
$M_W$ which is not susceptible to these saturation effects. For the
Harvard catalog, we study
the seismic moment distribution exclusively, and thereby also avoid
errors due to the discretization of the magnitude scale; our accuracy
is one order of magnitude greater than if we had used decimal
magnitudes [{\it Ekstrom and Dziewonski}, 1988]. {\it Hutton and Jones}
[1993] have
recently re-evaluated many of the events in the Southern California catalog and
have given
values of the moment magnitude $M_W$.

In this section, we perform a statistical analysis of the scalar
seismic moments $m_0$ in the Harvard catalog [{\it Dziewonski et al.}, 1993],
including
moment entries that have become available since the publication of {\it
Pacheco et
al.}\ [1992]. In contrast with our numerical tests above, we no longer
have the luxury of dealing with multiple realizations of the process:
we can only treat real catalogs as the result of a {\it single
realization}.  Fig. 4 presents the logarithm of the nth-ordered seismic
moment as a function of the logarithm of its rank, for the largest
shallow earthquakes (defined as those whose hypocentral depths are
less than $70$km), in the Harvard catalog. Two regimes can be clearly
distinguished: for the first $50$ largest earthquakes, we get $\mu_2 =
1.3 \pm0.1$, yielding $b_2 = 2.0 \pm0.2$ assuming $\beta = 1.5$, for
the large-energy branch. Here, the error bars are estimated from the
least-squares fit procedure and do not reflect the statistical error
inherent in a single realization. From the results obtained in section
3, in particular taking into account the existence of a skewness and
intrinsic statistical error, we correct this raw result and
write instead $\mu_2^{corrected} = 1.5 \pm0.2$ leading to  $b_2 = 2.3
\pm0.3$, a value to be compared with the earlier estimate  of
$b_2 = 1.6\pm0.1$ [{\it Pacheco et
al.}, 1992].  A value of $b_2$ near $2.3$ is wholly consistent
with the argument that the earthquake energy flux is finite
[{\it Knopoff and Kagan}, 1977;
{\it Kagan and Knopoff}, 1984;
{\it Kagan}, 1994];
however, it is quite inconsistent with the values suggested for either
a model wherein the moment of large ruptures is controlled by the
width of the rupture (W-model) or by its length (L-model)
derived from simple scaling arguments {\it Romanowicz and Rundle}, 1993].
A value of $b_2\simeq 2.3$ appears to be more compatible
with a W-model on the basis of a
recent calculation of the exponent by means of a
nonlinear diffusion equation [{\it Sornette and Sornette}, 1995]
which yields $b_2\simeq 3.$

For much larger ranks that correspond to the smaller earthquakes in
the catalog, we obtain  $b_1 =
1.00 \pm 0.02$ ($\mu = 0.67 \pm 0.01$) in rough agreement with
the value given by {\it Pacheco, et al.}\ [1992] $b_1 = 1.07 \pm 0.01$.
Note that this power law for
small events has a stable exponent only if we use very large ranks,
{from around $200$ and above}; we must avoid using data near the
apparent cross-over; the instability in the exponent always appears if
we have two regimes for the distribution (see Fig. 2).  It would thus
be erroneous to attempt to extract a second exponent for ranks between
say $50$ and $200$. Of course, the same conclusion is valid for the
cumulative distribution.  Using the results of
{\it Deemer and Votaw} [1955],
{\it Kagan} [1994] evaluated the
possibility that a distribution with two power-law exponents could approximate
the
Harvard data in the magnitude interval $ 5.8 \ge M_W \ge 7.8$, and found that
there is no statistically significant crossover at this magnitude
range.

{}From our tests on synthetic data sets, we have shown that the
determination of the cross-over value has a poor accuracy. It should
be stressed that this is true even when the particular realization
under study exhibits an apparent well-defined cross-over value, as is
the case for the Harvard earthquake catalog. Following {\it Pacheco et
al.}\ [1992], one could read on the plot the value $n_2 \simeq 50$ and
conclude therefore that there is a well-defined cross-over magnitude at $M_W
\simeq$  $7.5$, in agreement with the {\it Pacheco et al.}\ result of $M_W
\simeq$  $7.4$. We have noted
the danger in this identification when tested on synthetic data
sets: the existence of {\it intrinsic statistical fluctuations makes
the apparent cross-over fluctuate} within a factor two of the true
value from realization to realization. Thus our statistical tests have
shown that any claim of a good determination of a cross-over must be
tempered and that the true cross-over has a probability $1/2$ to be
between rank $30$ and rank $100$. Translated into magnitudes, there is
a probability of $1/2$ that the cross-over lies in the interval
$7.1<M_W<7.6$, an interval that corresponds to a down-dip thickness
$W$ of about 36 to 62 km,  values derived from the formula $W \sim (m_0/\Delta
\sigma)^{1/3}$ with a coefficient given by [{\it Pacheco et al.}, 1992];
$\Delta \sigma$ is the stress drop.

We point out that the data available to date are not sufficient in
numbers to certify the existence of a power law distribution for large
earthquakes.  We have found that a gamma distribution [{\it Kagan}, 1994]
also represents the data well and accurately. The gamma distribution
has only two adjustable parameters; a distribution with two power-law branches
has three adjustable parameters, and on grounds of statistical parsimony, a
data fitting scheme with a smaller number of parameters is to be preferred,
unless there is some compelling reason to use a more complex fitting scheme.
There is no doubt that there is a rollover or cross-over to a
distribution that falls off at a faster rate than for the smaller
earthquakes.  We have shown here that if we assume it to be a
cross-over to another power law, then we can estimate the value of its
exponent reasonably well in the case of the Harvard catalog; we cannot
estimate the cross-over moment or magnitude with any reasonable
certainty.

Fig. 5 shows the nth-ordered magnitude $M_W$ as a function of the
logarithm of its rank for the Southern California catalog [{\it Hutton and
Jones},
1993]. A single power law fits the whole range.  Since the magnitude (and
not the
logarithm of the moment) is plotted, the slope of the straight line
gives $\frac{1}{b}$ directly, yielding $b = 1.00 \pm 0.02$.  We thus
conclude that, in contrast with the results for the world-wide catalog
and previous claims [{\it Pacheco et al.}, 1992], the statistics for the
Southern
California catalog is fully compatible with a single power law distribution.
We
see no evidence for a bend or kink in the curve at $M= 5.9$ to
$6.0$.
This result causes us to question the validity of the proposal [{\it Pacheco et
al.}, 1992] that the thickness of the seismogenic zone determines the
cross-over
magnitude  for Southern California, and thus we may expect that
there is no observable
influence of source dimension on the frequency distribution of
earthquakes in other localities as well.

\section{The next big earthquake}

We comment that the rank-ordering technique may be used to
infer the size of the next forthcoming biggest event. If the
forthcoming large event is assumed to be imminent, its occurrence
corresponds to a  shift of the largest event recorded to date to rank
$2$, the second largest event to date to rank $3$, and so on. A fit of
the rank-ordering plot extrapolated to $n=1$ yields the seismic moment
of the next very largest earthquake. Performing this analysis on the
Southern California catalog, we obtain that the next largest event
should be of magnitude $M_W$ around $7.9 \pm 0.2$. The same analysis
applied to the Harvard catalog suggests that the next greatest
earthquake immediately larger than any in the catalog should have a
seismic moment $m_0 \simeq 6\cdot 10^{21} Nm$, i.e. $M_W \simeq 8.5
\pm 0.2$. The uncertainties have been estimated using eq.\ (3).

We can be more precise and use the maximum likelihood formalism
described above to infer the most probable value $E^{next}$ of the
energy of the next earthquake, restricted to be larger than some
threshold $E^*$ corresponding  the present rank $n$ (i.e. such that
$E^*=E_n$). In other words, we are discussing here {\it restricted
most probable values}. To do this calculation, we make use of eq.\ (4)
for the $n$th and $(n+1)$st events having
energies larger than $E^*=E_n$, assuming
that $n$ is sufficiently large, so that $1/\mu$ is close to its true
value. In this case $$1/\mu = \frac{1}{n} \sum_{i=1}^n \log
\frac{E_i}{E_n} = \frac{1}{n} \sum_{i=1}^{n+1} \log
\frac{E'_i}{E_n},$$ where the $E'_i=E_i$ for $E_i > E^{next}$ and
$E'_{i+1}=E_i$ for $E_i < E^{next}$. Thus the most
probable value of the next earthquake with $E$ larger than $E^*$ is
$E^{next} = E^* e^{\frac{1}{\mu}}$. This means roughly that
the next event will have the $(\frac{n}{e})$th rank. In other words, the
next event with $E$ larger than some threshold $E^*=E_n$ will not be,
in general, the largest of all, but only larger than the threshold by
a finite factor. If we stretch this argument to the unreal limit $n=1$
in the above formula, we can determine the most probable value of the
next earthquake, with $E$ larger than the largest observed to date,
$E_1$. We get $E^{next} = \frac{E_1}{E_2} E_1$ which is equal on
average to $E^{next} = (\frac{2\mu +1}{\mu +1})^{\frac{1}{\mu}}  E_1 =
2.3 E_1$ for $\mu=2/3$ corresponding to a magnitude gap of $0.6 \pm
0.3$; the uncertainty is estimated from eq.\ (3). We can ask a similar
question about the energy of the next earthquake such that its energy
will be
larger than the second largest earthquake observed to date. The result
is $E^{next} = (\frac{E_1}{E_2})^{1/2} E_2$. And so on.

\section{Summary}

The well-known rank-ordering statistical technique
is useful for extracting the tail of the distribution for a
sparse data set. The combination of analytical results and numerical
tests has allowed us to quantify the accuracy that can be obtained
for the determination of the exponents of power law distributions. We
have confirmed that two power law branches to the distribution might
account for the moments in the Harvard seismic moment catalog and we
have corrected our estimate of the two corresponding $b$-values for
skewness and intrinsic statistical errors. However, the cross-over
value between the two power laws has been shown to be ill-defined. In
the case of the Southern California catalog, we have shown that a
single power law distribution, without any large-energy branch
satisfies the observations; we find no evidence for a bend or kink in the
distribution, nor can we make any inference about the thickness of the
seismogenic zone  from the earthquake size distribution.

We believe that this technique will be useful to analyze other
geological data sets, often characterized by undersampled fat tails
that correspond to rare extreme events.

\section{Acknowledgments}

D.S. acknowledges stimulating discussions with J.-P. Bouchaud. This
work has been partially supported by the CNRS-NSF International
Cooperation program, and by the Southern California Earthquake Center.
Publication No. XXXX, Institute of Geophysics and Planetary Physics,
University of California, Los Angeles.  Publication No. YYYY, Southern
California Earthquake Center.

\vskip 1truecm
\pagebreak
{\Large \bf References}

\begin{itemize}

\item Aharony A. and J. Feder eds., {\it Fractals in Physics},
North Holland, Amsterdam, Physica D, {\bf 38}, nos. 1-3, 1989.

\item Aki K., Maximum likelihood estimate of $b$ in the formula
$\log N = a - bm$ and its confidence limits, Bull. Earthquake Res. Inst. Tokyo
Univ., 43, 237-239 (1965).

\item Deemer W.L. and D.F. Votaw,
Estimation of parameters of truncated or censored exponential
distributions,
Ann. Math. Stat., {\bf 26}, 498-504, 1955.

\item Dziewonski A.M., G. Ekstrom, and M. P. Salganik,
Centroid-moment tensor solutions for January-March, 1992,
Phys. Earth Planet. Inter., {\bf 77}, 143-150, 1993, and references
therein.

\item Ekstrom G. and A. M. Dziewonski,
Evidence of bias in estimation of earthquake size,
Nature, {\bf 332}, 319-323, 1988.

\item Epstein B. and M. Sobel, Life testing, J. Amer. Statist.
Assoc., {\bf 48}, 486-502, 1953.

\item  Geller R.J.,
Scaling relations for earthquake source parameters and
magnitudes.
Bull. Seismol. Soc. Amer., {\bf 66}, 1501-1523, 1976.

\item Gumbel E.J., {\it Statistics of Extremes,} Columbia Univ.
Press, New York, 1960.

\item Hill, B.M. A simple general approach to inference about
the tail of a distribution, Ann. Statistics, {\bf 3}, 1163-1174, 1975.

\item Hutton L.K. and L.M. Jones, Local magnitudes and apparent
variations in seismicity rates in Southern California, Bull. Seismol.
Soc. Amer., {\bf 83}, 313-329, 1993, and references therein.

%
\item Kagan Y.Y.,
Observational evidence for earthquakes as a nonlinear dynamic
process, Physica D, {\bf 77}, 160-192, 1994. Also appeared in
{\it Modeling the Forces of Nature}, eds. R. Camassa,
J. M. Hyman, and W. I. Newman, North-Holland, Amsterdam,
160-192, 1994.

\item Kagan Y.Y. and L. Knopoff,
A stochastic model of earthquake occurrence,
Proc. 8th Int. Conf. Earthq. Eng., {\bf 1}, 295-302, 1984.

\item Kanamori H.,
The energy release in great earthquakes,
J. Geophys. Res., {\bf 82}, 2981-2987, 1977.

\item Kanamori H. and D.L.  Anderson,
Theoretical basis of some empirical relations in seismology,
Bull. Seism. Soc. Am., {\bf 65}, 1073-1095, 1975.

\item Knopoff L. and Y. Kagan, Analysis of the theory of extremes as
applied to earthquake
problems,
J. Geophys. Res. {\bf 82}, 5647-5657, 1977.

\item Mandelbrot B.B., {\it The Fractal Geometry of Nature},
Freeman, New York, 1983.

\item Okal E.A. and B.A. Romanowicz,  On the variation of
$b$-values with earthquake size.  Phys. Earth Planet. Inter., {\bf 87},
55-76, 1994.

\item Pacheco J.F., C.H. Scholz and L.R. Sykes,
Changes in frequency-size relationship from small to large
earthquakes,
Nature, {\bf 355}, 71-73, 1992.

\item Riste T. and D. Sherrington eds.,
{\it Spontaneous formation of space-time structures and criticality},
Proc. NATO ASI, (Geilo, Norway),
Kluwer, Dordrecht, 1991.

\item Romanowicz B.,
Strike-slip earthquakes on quasi-vertical transcurrent
faults: inferences for general scaling behavior,
Geophys. Res. Lett., {\bf 19}, 481-484, 1992.

\item Romanowicz B. and J.B. Rundle, On scaling relations for
large earthquakes, Bull. Seism. Soc. Am., {\bf 83},
1294-1297, 1993.

\item Rundle J.B.,
Derivation of the complete Gutenberg-Richter
magnitude-frequency relation using the principle of scale
invariance,
J. Geophys. Res., {\bf 94}, 12,337-12,342, 1989.

\item Scholz C., {\it The mechanics of Earthquakes and
Faulting}, Cambridge
University Press, Cambridge, 1990.

\item Shimazaki K., Small and large earthquakes:
The effects of the thickness of
seismogenic layer and the free surface. In {\it Earthquake Source
Mechanics}, S. Das, J. Boatwright and C. Scholz, eds.,
Geophys. Monograph {\bf37}, Amer. Geophys. Un.,Washington, 1986, pp 209-216.

\item Sornette D. and A. Sornette,  Theoretical implications
of a recent non-linear diffusion equation linking short-time
deformation to long-time tectonics, Bull. Seism. Soc. Am., in press, 1995.

\item Zipf G.K., {\it Human Behavior and the Principle of
Least-effort}, Addison-Wesley, Cambridge, 1949.

\end{itemize}

\pagebreak

FIGURE CAPTIONS:
\vskip 1truecm

Fig. 1. Rank-order distribution for five
different realizations of a random process whose values are distributed
according
to the power law (1) with $\mu = 1$. $N= 10^5$ events were generated
numerically
for each realization and the $100$ largest events were retained to
construct the distributions.
 The two straight lines represent the expected standard deviation interval
obtained
from (3).

\vskip 0.5 truecm

Fig. 2  Rank-ordering  of a single simulation run with parameters
$N = 10^5$, $\mu_1 = \frac{2}{3}$, $\mu_2 = \frac{4}{3}$ and
$n_2 = 50$, corresponding to $E_c \sim 10^5$.
The two power law branches of the distribution are clearly
identifiable.

\vskip 0.5 truecm

Fig. 3   Histogram of
the number of  realizations with the
same  $\mu_{2,est}$ out of $10^5$ realizations.
Each realization was generated to have $N = 10^5$ events,
$\mu_1 = \frac{2}{3}$, $\mu_2 = \frac{4}{3}$, and
cross-over value $n_2 = 50$   (i.e. $E_c \sim 10^5$).
The five curves correspond to different values of $n$ ($n= 25,
50, 75, 100$ and $200$) which is the interval over which
the estimate of the exponent $\mu_{2,est}$ is extracted.

\vskip 0.5 truecm

Fig. 4   Log-log plot of the rank-ordered seismic moment (in units of $10^{19}
Nm$) of the largest shallow earthquakes ($h< 70$km) in the Harvard catalog
versus
its rank.

\vskip 0.5 truecm

Fig. 5  The rank-ordered magnitude $M_W$ of the strongest earthquakes
in the Southern California catalog vs. the logarithm of its rank.

\end{document}